\documentclass[11pt]{article}

\usepackage{xspace}
\usepackage{theorem}
\usepackage{amsmath,amsfonts,amssymb}
\usepackage{pstricks,pst-node}
\usepackage{verbatim}
\usepackage{graphicx}

\newcommand{\Calig}[1]{%                calligraphic text
  \ensuremath{\mathcal{#1}}}

\newcommand{\DL}{DL\xspace}
\newcommand{\KB}{KB\xspace}

\newcommand{\ALC}{\Calig{ALC}\xspace}
\newcommand{\Galen}{\textsc{Galen}\xspace}
\newcommand{\Fact}{FaCT\xspace}

\newcommand{\Kris}{\textsc{Kris}\xspace}

\newcommand{\I}{\Calig{I}\xspace}
\newcommand{\Ifunc}{^\I}
\newcommand{\Domain}{\ensuremath{\Delta\Ifunc}\xspace}
\newcommand{\Doti}{\ensuremath{\cdot\Ifunc}\xspace}
\newcommand{\Tau}{\Calig{T}\xspace}

\newcommand{\Not}{\ensuremath{\neg}}
\newcommand{\Some}[2]{%
  \ensuremath{\exists #1.#2}}
\newcommand{\All}[2]{%
  \ensuremath{\forall #1.#2}}

\newcommand{\Cname}[1]{\mbox{\small \textsf{#1}}}
\newcommand{\Rname}[1]{\mbox{\small \textsf{\textsl{#1}}}}

\newcommand{\Lab}{\Calig{L}}

\newcommand{\Implies}{\ensuremath{\Rightarrow}}
\newcommand{\Iff}{\ensuremath{\mathbin{\mathsf{iff}}}}

\newfont{\bigmathxx}{cmsy10 scaled 2000}

\theoremstyle{plain}
\theoremheaderfont{\bf\boldmath}
\newtheorem{satz}{Satz}[section]
\newtheorem{theorem}[satz]{Theorem}
\newtheorem{definition}[satz]{Definition}
\theoremstyle{plain}
\newtheorem{lemma}[satz]{Lemma}

\newcommand{\cW}{\Calig{W}\xspace}
\newcommand{\sL}{\ensuremath{\mathsf{L}}\xspace}

\newcommand{\cT}{\Calig{T}\xspace}

\newcommand{\Int}{\ensuremath{\mathsf{Int}}}
\newcommand{\NC}{\ensuremath{\mathsf{NC}}\xspace}
\newcommand{\NR}{\ensuremath{\mathsf{NR}}\xspace}

\newenvironment{proof}{\noindent \textit{Proof.}}{}

\newcommand{\qed}{\hfill \vrule height4pt width 4pt depth0pt}

\newcommand{\witn}[1]{(\textsf W#1)}

\newcommand{\interp}[1]{(\textsf I#1)}

\newcommand{\wrt}{w.r.t.\ }

\newcommand{\LabW}{\ensuremath{\Lab^\cW}\xspace}
\newcommand{\Tu}{\ensuremath{\mathcal{T}_u}\xspace}
\newcommand{\Tg}{\ensuremath{\mathcal{T}_g}\xspace}

\title{Reasoning with Axioms: Theory and Practice\thanks{This paper appeared
    in the Proceedings of the Seventh International Conference on Priciples of 
    Knowledge Representation and Reasoning (KR'2000).}}
\author{
  {\bf Ian Horrocks}\\
  Department of Computer Science\\
  University of Manchester, UK\\
  {\tt horrocks@cs.man.ac.uk} \and
  {\bf Stephan Tobies}\\
  LuFg Theoretical Computer Science\\ 
  RWTH Aachen, Germany \\
  {\tt tobies@informatik.rwth-aachen.de}}
\date{}

\begin{document}

\maketitle

\begin{abstract}
  When reasoning in description, modal or temporal logics it is often useful
  to consider axioms representing universal truths in the domain of discourse.
  Reasoning with respect to an arbitrary set of axioms is hard, even for
  relatively inexpressive logics, and it is essential to deal with such axioms
  in an efficient manner if implemented systems are to be effective in real
  applications. This is particularly relevant to Description Logics, where
  subsumption reasoning with respect to a terminology is a fundamental
  problem. Two optimisation techniques that have proved to be particularly
  effective in dealing with terminologies are lazy unfolding and absorption.
  In this paper we seek to improve our theoretical understanding of these
  important techniques. We define a formal framework that allows the
  techniques to be precisely described, establish conditions under which they
  can be safely applied, and prove that, provided these conditions are
  respected, subsumption testing algorithms will still function correctly.
  These results are used to show that the procedures used in the \Fact system
  are correct and, moreover, to show how efficiency can be significantly
  improved, while still retaining the guarantee of correctness, by relaxing
  the safety conditions for absorption.

\end{abstract}

\section{MOTIVATION}\label{sec:motivat}

Description Logics (DLs) form a family of formalisms which have grown
out of knowledge representation techniques using frames and semantic
networks. DLs use a class based paradigm, describing the domain of
interest in terms of concepts (classes) and roles (binary relations)
which can be combined using a range of operators to form more complex
structured concepts~\cite{Baader91b}.  A DL \emph{terminology}
typically consists of a set of asserted facts, in particular asserted
subsumption (is-a-kind-of) relationships between (possibly complex) concepts.\footnote{DLs can
  also deal with assertions about individuals, but in this paper we
  will only be concerned with \emph{terminological} (concept based)
  reasoning}.

One of the distinguishing characteristics of DLs is a formally defined
semantics which allows the structured objects they describe to be
reasoned with.  Of particular interest is the computation of implied
subsumption relationships between concepts, based on the assertions in
the terminology, and the maintenance of a concept hierarchy (partial
ordering) based on the subsumption relationship~\cite{Woods92}.

The problem of computing concept subsumption relationships has been
the subject of much research, and sound and complete algorithms are
now known for a wide range of DLs (for
example~\cite{Hollunder90b,Baader91e,Baader91c,DeGiacomo98a,Horrocks99j}).
However, in spite of the fundamental importance of terminologies in
DLs, most of these algorithms deal only with the problem of deciding
subsumption between two concepts (or, equivalently, concept
satisfiability), without reference to a terminology (but
see~\cite{Buchheit93,Calvanese96a,Donini96a,Horrocks99j}).  
By restricting the kinds of assertion that can appear in a terminology,
concepts can be syntactically expanded so as to explicitly include all
relevant terminological information. This procedure, called \emph{unfolding},
has mostly been applied to less expressive DLs.
 With more expressive DLs, in particular those
supporting universal roles, it is often possible to encapsulate an
arbitrary terminology in a single concept. This technique can be used
with satisfiability testing to ensure that the result is valid with
respect to the assertions in the terminology, a procedure called
\emph{internalisation}.

Although the above mentioned techniques suffice to demonstrate the
theoretical adequacy of satisfiability decision procedures for
terminological reasoning, experiments with implementations have shown
that, for reasons of (lack of) efficiency, they are highly
unsatisfactory as a practical methodology for reasoning with DL
terminologies. Firstly, experiments with the \Kris system have shown
that integrating unfolding with the (tableaux) satisfiability algorithm
(\emph{lazy unfolding}) leads to a significant improvement in
performance~\cite{BFHNP94}. More recently, experiments with the
\Fact system have shown that reasoning becomes hopelessly intractable
when internalisation is used to deal with larger
terminologies~\cite{Horrocks98c}.  However, the \Fact system has also
demonstrated that this problem can be dealt with (at least for
realistic terminologies) by using a combination of lazy unfolding and
internalisation, having first manipulated the terminology in order to
minimise the number of assertions that must be dealt with by
internalisation (a technique called \emph{absorption}).

It should be noted that, although these techniques were discovered
while developing DL systems, they are applicable to a whole range of
reasoning systems, independent of the concrete logic and type of algorithm.
As well as tableaux based decision procedures, this includes
resolution based algorithms, where the importance of minimising the
number of terminological sentences has already been
noted~\cite{Hustadt99a}, and sequent calculus algorithms, where there
is a direct correspondence with tableaux algorithms~\cite{Horrocks99i}.

In this paper we seek to improve our theoretical understanding of
these important techniques which has, until now, been very limited.
In particular we would like to know exactly when and how they can be
applied, and be sure that the answers we get from the algorithm are
still correct. This is achieved by defining a formal framework that
allows the techniques to be precisely described, establishing
conditions under which they can be safely applied, and proving that,
provided these conditions are respected, satisfiability algorithms
will still function correctly.
These results are then used to show that the procedures used in the
\Fact system are correct\footnote{Previously, the correctness of these 
  procedures had only been demonstrated by a relatively ad-hoc
  argument~\cite{Horrocks97b}.} and, moreover, to show how efficiency can
be significantly improved, while still retaining the guarantee of
correctness, by relaxing the safety conditions for absorption.
Finally, we identify several interesting directions for future
research, in particular the problem of finding the ``best'' absorption
possible.

\section{PRELIMINARIES}

Firstly, we will establish some basic definitions that clarify what we mean by
a \DL, a terminology (subsequently called a TBox), and subsumption and
satisfiability with respect to a terminology, . The results in this paper are
uniformly applicable to a whole range of \DL{}s, as long as some basic
criteria are met:

\begin{definition}[Description Logic]
  Let \sL be a \DL based on infinite sets of atomic concepts \NC and atomic
  roles \NR.  We will identify \sL with the sets of its well-formed concepts
  and require \sL to be closed under boolean operations and sub-concepts. 
  
  An interpretation is a pair $\I = (\Domain, \Doti)$, where \Domain is a
  non-empty set, called the \emph{domain} of $\I$, and \Doti is a function mapping \NC
  to $2^{\Domain}$ and \NR to $2^{\Domain \times \Domain}$. With each DL \sL
  we associate a set $\Int(\sL)$ of \emph{admissible} interpretations for \sL.
  $\Int(\sL)$ must be closed under isomorphisms, and, for any two
  interpretations $\I$ and $\I'$ that agree on \NR, it must satisfy $\I \in
  \Int(\sL) \Leftrightarrow \I' \in \Int(\sL)$. Additionally, we assume that
  each \DL \sL comes with a semantics that allows any interpretation $\I \in
  \Int(\sL)$ to be extended to each concept $C \in \sL$ such that it satisfies
  the following conditions:
  \begin{itemize}
  \item[\interp 1] it maps the boolean combination of concepts to the
    corresponding boolean combination of their interpretations, and
  \item [\interp 2] the interpretation $C^\I$of a compound concept $C \in \sL$
    depends only on the interpretation of those atomic concepts and roles that
    appear syntactically in $C$.
  \end{itemize}
\end{definition}

This definition captures a whole range of \DL{}s, namely, the important DL
\ALC~\cite{SSSm91} and its many extensions. $\Int(\sL)$ hides restrictions on
the interpretation of certain roles like transitivity, functionality, or role
hierarchies, which are imposed by more expressive \DL{}s (e.g.,
\cite{Horrocks99j}), as these are irrelevant for our purposes. In these cases,
$\Int(\sL)$ will only contain those interpretations which interpret the roles
as required by the semantics of the logic, e.g., features by partial functions 
or transitively closed roles by transitive relations. Please note that various modal
logics \cite{Schi91}, propositional dynamic logics \cite{DeLe94} and temporal
logics~\cite{Emerson85} also fit into this framework. We will use $C
\rightarrow D$ as an abbreviation for $\neg C \sqcup D$, $C \leftrightarrow D$
as an abbreviation for $(C \rightarrow D) \sqcap (D \rightarrow C)$, and
$\top$ as a tautological concept, e.g., $A \sqcup \neg A$ for an arbitrary $A
\in \NC$.

A TBox consists of a set of axioms asserting
subsumption or equality relations between (possibly complex) concepts.

\begin{definition}[TBox, Satisfiability]
  A \emph{TBox} \cT for \sL is a finite set of axioms of the form $C_1
  \sqsubseteq C_2$ or $C_1 \doteq C_2$, where $C_i \in \sL$. If, for some $A
  \in \NC$, \Tau contains one or more axioms of the form $A \sqsubseteq C$ or
  $A \doteq C$, then we say that $A$ is \emph{defined} in \Tau.

  Let $\sL$ be a DL and \cT a TBox. An interpretation $\I \in
  \Int(\sL)$ is a \emph{model} of \cT iff, for each $C_1 \sqsubseteq C_2 \in
  \cT$, $C_1\Ifunc \subseteq C_2\Ifunc$ holds, and, for each $C_1 \doteq C_2
  \in \cT$, $C_1\Ifunc = C_2\Ifunc$ holds. In this case we write $\I \models
  \cT$.  A concept $C \in \sL$ is \emph{satisfiable} with respect to a TBox
  \cT iff there is an $\I \in \Int(\sL)$ with $\I \models \cT$ and $C\Ifunc
  \neq \emptyset$. 
  A concept $C \in \sL$ \emph{subsumes} a
  concept $D \in \sL$ \wrt \cT iff, for all $\I \in \Int(\sL)$ with $\I
  \models \cT$, $C\Ifunc \supseteq D\Ifunc$ holds.

  Two TBoxes $\cT, \cT'$ are called \emph{equivalent} ($\cT \equiv
  \cT')$, iff, for all $\I \in \Int(\sL)$, $\I \models \cT \Iff \I \models \cT'$.
\end{definition}

We will only deal with concept satisfiability as concept subsumption can be
reduced to it for \DL{}s that are closed under boolean operations: $C$ subsumes
$D$ \wrt \cT iff $(D \sqcap \Not C)$ is not satisfiable \wrt \cT.

For temporal or modal logics, satisfiability with respect to a set of formulae
$\{C_1, \dots, C_k\}$ asserted to be universally true corresponds to
satisfiability \wrt the TBox $\{\top \doteq C_1, \dots, \top \doteq C_n \}$.

Many decision procedures for \DL{}s base their judgement on the existence of
models or pseudo-models for concepts. A central r\^ole in these algorithms is played
by a structure that we will call a \emph{witness} in this paper. It generalises the
notions of \emph{tableaux} that appear in \DL tableau-algorithms%
~\cite{HoNS90,BaaderBuchheit+-AIJ-1996,Horrocks99j} as well as the
\emph{Hintikka-structures} that are used in tableau and automata-based
decision procedures for temporal logic \cite{Emerson85} and propositional
dynamic logic~\cite{VaWo86}.

\begin{definition}[Witness]\label{def:witness}
  Let \sL be a \DL and $C \in \sL$ a concept. A \emph{witness}
  $\cW = (\Delta^\cW, \cdot^\cW, \LabW)$ for $C$ consists of a
  non-empty set $\Delta^\cW$, a function $\cdot^\cW$ that maps \NR to
  $2^{\Delta^{\cW} \times \Delta^{\cW}}$, and a function \LabW that
  maps $\Delta^\cW$ to $2^{\sL}$ such that the following properties
  are satisfied:
  \begin{itemize}
  \item[\witn 1] there is some $x \in \Delta^\cW$ with $C \in \LabW(x)$,
  \item[\witn 2] there is an interpretation $\I \in \Int(\sL)$ that
    \emph{stems} from \cW, and
  \item [\witn 3] for each interpretation $\I \in \Int(\sL)$
    that \emph{stems} from \cW, it holds that $D \in \LabW(x)$ implies
    $x \in D\Ifunc$.
  \end{itemize}
  An interpretation $\I=(\Domain,\Doti)$ is said to stem from \cW if it
  satisfies:
  \begin{enumerate}
  \item $\Domain = \Delta^\cW$,
  \item $\Doti|_{\NR} = \cdot^\cW$, and
  \item for each $A \in \NC$, $A \in \LabW(x) \ \Implies \ x \in
    A\Ifunc$ and $\neg A \in \LabW(x) \ \Implies \ x \not\in A\Ifunc$.
  \end{enumerate}
  
  A witness \cW is called \emph{admissible} with respect to
  a TBox \cT if there is an interpretation $\I \in \Int(\sL)$ that stems from
  \cW with $\I \models \cT$.
\end{definition}

Please note that, for any witness \cW, \witn 2 together with Condition 3 of
``stemming'' implies that, there exists no $x \in \Delta^\cW$ and $A \in \NC$,
such that $\{A, \neg A\} \subseteq \LabW(x)$.  Also note that, in general,
more than one interpretation may stem from a witness. This is the case if, for
an atomic concept $A \in \NC$ and an element $x \in \Delta^\cW$, $\Lab^\cW(x)
\cap \{ A, \neg A \} = \emptyset$ holds (because two interpretations
$\I$ and $\I'$, with $x \in A^\I$ and $x \in \neg A^{\I'}$, could both
stem from \cW).

Obviously, each interpretation $\I$ gives rise to a special witness, called
the \emph{canonical witness}:

\begin{definition}[Canonical Witness]
  Let \sL be a \DL. For any interpretation $\I \in \Int(\sL)$ we define the
  \emph{canonical witness} $\cW_\I = (\Delta^{\cW_\I}, \cdot^{\cW_\I},
  \Lab^{\cW_\I})$ as follows:
  \begin{align*}
    \Delta^{\cW_\I} & = \Domain \\
    \cdot^{\cW_\I} & = \Doti|_{\NR} \\
    \Lab^{\cW_\I} & = \lambda x.\{ D \in \sL \mid x \in D\Ifunc \}
  \end{align*}
\end{definition}

The following elementary properties of a canonical witness will be useful in
our considerations.

\begin{lemma}\label{lem:canonical-witness}
  Let \sL be a \DL, $C \in \sL$, and \cT a TBox. For each $\I \in
  \Int(\sL)$ with $C\Ifunc \neq \emptyset$, 
  \begin{enumerate}
  \item each interpretation $\I'$ stemming from $\cW_\I$ is isomorphic to \I
  \item $\cW_\I$ is a witness for $C$,
  \item $\cW_\I$ is admissible \wrt \cT iff $\I \models \cT$
  \end{enumerate}
\end{lemma}

\begin{proof}
  \begin{enumerate}
  \item Let $\I'$ stem from $\cW_\I$. This implies $\Delta^{\I'} = \Domain$
    and $\cdot^{\I'}|_{\NR} = \Doti|_{\NR}$. For each $x \in
    \Domain$ and $A \in \NC$, $\{A, \neg A\} \cap \Lab^{\cW_\I}(x) \neq
    \emptyset$, this implies $\cdot^{\I'}|_{\NC} = \Doti|_{\NC}$ and hence \I and 
    $\I'$ are isomorphic.
  \item Properties \witn 1 and \witn 2 hold by construction. Obviously, \I stems 
    from $\cW_\I$ and from (1) it follows that each interpretation $\I'$
    stemming from $\cW_\I$ is isomorphic to \I, hence \witn 3 holds.
  \item Since \I stems from $\cW_\I$, $\I \models \cT$ implies that $\cW_\I$
    is admissible \wrt \cT. If $\cW_\I$ is admissible \wrt \cT, then there is an
    interpretation $\I'$ stemming from $\cW_\I$ with $\I' \models \cT$. Since
    $\I$ is isomorphic to $\I'$, this implies $\I \models \cT$. \qed
  \end{enumerate}
\end{proof}

As a corollary we get that the existence of admissible witnesses is closely
related to the satisfiability of concepts \wrt TBoxes:

\begin{lemma}\label{cor:witness-and-satisfiability}
  Let \sL be a \DL. A concept $C \in \sL$ is satisfiable \wrt a TBox \cT iff
  it has a witness that is admissible \wrt \cT.
\end{lemma}

\begin{proof}
  For the \emph{only if}-direction let $\I \in \Int(\sL)$ be an interpretation
  with $\I \models \cT$ and $C\Ifunc \neq \emptyset$. From Lemma~\ref{lem:canonical-witness} it follows that
  the canonical witness $\cW_\I$ is a witness for $C$ that is admissible
  \wrt \cT.
  
  For the \emph{if}-direction let \cW be an witness for $C$ that
  is admissible \wrt \cT. This
  implies that there is an interpretation $\I \in \Int(\sL)$ stemming from \cW
  with $\I \models \cT$. For each interpretation \I that stems from \cW, it
  holds that $C\Ifunc \neq \emptyset$ due to \witn 1 and \witn 3. \qed
\end{proof}

From this it follows that one can test the satisfiability of a concept \wrt to 
a TBox by checking for the existence of an admissible witness. We call
algorithms that utilise this approach \emph{model-building algorithms}.

This notion captures tableau-based decision procedures,
\cite{HoNS90,BaaderBuchheit+-AIJ-1996,Horrocks99j}, those using
automata-theoretic approaches \cite{VaWo86,CaDL99} and, due to their direct
correspondence with tableaux algorithms~\cite{Hustadt99a,Horrocks99i}, even
resolution based and sequent calculus algorithms.

 %This work develops a

The way many decision procedures for \DL{}s deal with TBoxes exploits the
following simple lemma.

\begin{lemma}\label{lem:simple-approach}
  Let \sL be a \DL, $C \in \sL$ a concept, and \cT a TBox. Let \cW be a
  witness for $C$. If
  %, for each $x \in \Delta^\cW$,
  \[
   \begin{array}{lcl}
    C_1 \sqsubseteq C_2 \in \cT  & \Implies & \forall x \in \Delta^\cW.(
    C_1  \rightarrow C_2 \in \LabW(x))\\
    C_1 \doteq C_2 \in \cT \; & \Implies & \forall x \in \Delta^\cW.(C_1 \leftrightarrow C_2 \in \LabW(x))
  \end{array}
  \]
  then \cW is admissible \wrt \cT.
\end{lemma}

\begin{proof}
  \cW is a witness, hence there is an interpretation $\I \in \Int(\sL)$
  stemming from \cW. From \witn 3 and the fact that \cW satisfies the
  properties stated in \ref{lem:simple-approach} it
  follows that, for each $x \in \Domain$,
  \[
  \begin{array}{lcl}
    C_1 \sqsubseteq C_2 \in \cT  & \Implies &  C_1 \rightarrow C_2 \in
    \LabW(x) \\ & \Implies & x \in (C_1 \rightarrow C_2)\Ifunc\\
    C_1 \doteq C_2 \in \cT  & \Implies & C_1 \leftrightarrow C_2 \in \LabW(x)
    \\ 
    & \Implies & x \in (C_1 \leftrightarrow C_2)\Ifunc 
  \end{array}
  \]
  Hence, $\I \models \cT$ and \cW is admissible \wrt \cT. \qed
\end{proof}

Examples of algorithms that exploit this lemma to deal with axioms can be
found in \cite{Donini96a,DeLe96,Horrocks99j}, where, for each axiom $C_1
\sqsubseteq C_2$ ($C_1 \doteq C_2$) the concept $C_1 \rightarrow C_2$ ($C_1
\leftrightarrow C_2$) is added to every node of the generated tableau. 

Dealing with general axioms in this manner is costly due to the high degree of
nondeterminism introduced. This can best be understood by looking at tableaux
algorithms, which try to build witnesses in an incremental fashion. For a
concept $C$ to be tested for satisfiability, they start with $\Delta^\cW = \{
x_0 \}$, $\LabW(x_0) = \{ C \}$ and $\cdot^\cW(R) = \emptyset$ for each $R \in
\NR$. Subsequently, the concepts in $\LabW$ are decomposed and, if necessary,
new nodes are added to $\Delta^\cW$, until either \cW is a
witness for $C$, or an obvious contradiction of the form $\{A,\neg A\}
\subseteq \LabW(x)$, which violates \witn 2, is
generated. In the latter case, backtracking search is used to explore
alternative non-deterministic decompositions (e.g., of disjunctions),
one of which could lead to the discovery of a witness.

When applying Lemma~\ref{lem:simple-approach}, disjunctions are added
to the label of each node of the tableau for each general axiom in the
TBox (one disjunction for axioms of the form $C_1 \sqsubseteq C_2$,
two for axioms of the form $C_1 \doteq C_2$). This leads to an
exponential increase in the search space as the number of nodes and
axioms increases. For example, with 10 nodes and a TBox containing 10
general axioms (of the form $C_1 \sqsubseteq C_2$) there are already
100 disjunctions, and they can be non-deterministically decomposed in
$2^{100}$ different ways. For a TBox containing large numbers of
general axioms (there are 1,214 in the \Galen medical terminology
\KB~\cite{Rector93a}), this can degrade performance to the extent that
subsumption testing is effectively non-terminating. To reason with
this kind of TBox we must find a more efficient way to deal with
axioms.

\section{ABSORPTIONS}

We start our considerations with an analysis of a technique
that can be used to deal more efficiently with so-called
primitive or acyclic TBoxes.

\begin{definition}[Absorption]
  Let \sL be a \DL and \cT a TBox. An \emph{absorption} of \cT is a pair of
  TBoxes $(\Tu,\Tg)$ such that $\cT \equiv \Tu \cup \Tg$ and \Tu contains only
  axioms of the form $A \sqsubseteq D$ and $\neg A \sqsubseteq D$ where $A \in
  \NC$.% is an atomic concept.

  An absorption $(\Tu,\Tg)$ of \cT is called \emph{correct} if it satisfies
  the following condition. For each witness \cW, if, for each $x \in  \Delta^\cW$,
  \[
    \begin{array}{r@{\;}c@{\;}l}
      A \sqsubseteq D \in \Tu \wedge  A \in \LabW(x) & \Implies& D \in
      \LabW(x)\\
      \neg A \sqsubseteq D \in \Tu  \wedge \neg A \in \LabW(x) & \Implies & D \in
      \LabW(x)\\
      C_1 \sqsubseteq C_2 \in \Tg & \Implies & C_1 \rightarrow C_2 \in
      \LabW(x)\\
      C_1 \doteq C_2 \in \Tg & \Implies & C_1 \leftrightarrow C_2 \in \LabW(x)
    \end{array}
  \]
  then \cW is admissible \wrt \cT. We refer to this properties by $(*)$. A
  witness that satisfies $(*)$ will be called \emph{unfolded \wrt \cT{}}.
\end{definition}

If the reference to a specific TBox is clear from the context, we will often
leave the TBox implicit and say that a witness is unfolded.

How does a correct absorption enable an algorithm to deal with axioms more
efficiently? This is best described by returning to tableaux
algorithms. Instead of dealing with axioms as previously described, which may
lead to an exponential increase in the search space, axioms in $\Tu$ can now
be dealt with in a deterministic manner. Assume, for example, that we have to handle the axiom 
$A \doteq C$. If the label of a node already
contains $A$ (resp.\ $\neg A$), then $C$ (resp.\ $\neg C$) is added to
the label; if the label contains neither $A$ nor $\neg A$, then
\emph{nothing} has to be done. Dealing with the
axioms in $\Tu$ this way avoids the necessity for additional non-deterministic
choices and leads to a gain in efficiency. A witness produced in this manner will be
unfolded and is a certificate for satisfiability \wrt \cT.  This technique is generally known
as \emph{lazy unfolding} of primitive TBoxes~\cite{Horrocks98c}; formally, it is
justified by the following lemma:

\begin{lemma}
  Let $(\Tu,\Tg)$ be a correct absorption of \cT. For any $C \in \sL$, $C$ has 
  a witness that is admissible \wrt \cT iff $C$ has an unfolded witness.
\end{lemma}

\begin{proof}
  The \emph{if}-direction follows from the definition of ``correct
  absorption''. For the \emph{only if}-direction, let $C \in \sL$ be a concept
  and \cW a witness for $C$ that is admissible \wrt \cT. This implies the
  existence of an interpretation $\I \in \Int(\sL)$ stemming from \cW such
  that $\I \models \cT$ and $C^\I \neq \emptyset$. Since $\cT \equiv \Tu \cup
  \Tg$ we have $\I \models \Tu \cup \Tg$ and hence the canonical witness
  $\cW_\I$ is an unfolded witness for $C$. \qed
\end{proof}

A family of TBoxes where absorption can successfully be applied are
\emph{primitive} TBoxes, the most simple form of TBox usually studied in the
literature.

\begin{definition}[Primitive TBox]\label{def:primitive-tbox}
  A TBox \cT is called \emph{primitive} iff it consists entirely of axioms of the form
  $A \doteq D$ with $A \in \NC$, each $A \in \NC$ appears as at most one
  left-hand side of an axiom, and \cT is acyclic. Acyclicity is defined as
  follows: $A \in \NC$ is said to \emph{directly use} $B \in \NC$ if $A \doteq
  D \in \cT$ and $B$ occurs in $D$; \emph{uses} is the transitive
  closure of ``directly uses''. We say that \cT is \emph{acyclic} if there is
  no $A \in \NC$ that uses itself.
\end{definition}

For primitive TBoxes a correct absorption can easily be given.

\begin{theorem}\label{theo:prim-tbox}
  Let \cT be a primitive TBox, $\Tg = \emptyset$, and \Tu defined by 
  \[
  \Tu = \{ A \sqsubseteq D, \neg A \sqsubseteq \neg D \mid A \doteq D \in \cT
  \} .
  \] %
  Then $(\Tu,\Tg)$ is a correct absorption of \cT.
\end{theorem}

\begin{proof}
  Trivially, $\cT \equiv \Tu \cup \Tg$ holds. Given an unfolded witness $\cW$,
  % = (\Delta^\cW,\cdot^\cW,\LabW)$
  we have to show that
  there is an interpretation \I stemming from \cW with $\I \models \cT$.  
  
  We fix an arbitrary linearisation $A_1, \dots, A_k$ of the ``uses'' partial
  order on the atomic concept names appearing on the left-hand sides of axioms
  in \cT such that, if $A_i$ uses $A_j$, then $j < i$ and the defining concept
  for $A_i$ is $D_i$.
  
  For some interpretation \I, atomic concept $A$, and set $X \subseteq
  \Delta^\I$, we denote the interpretation that maps $A$ to $X$ and agrees
  with $\I$ on all other atomic concepts and roles by $\I[A \mapsto X]$.  For
  $0 \leq i \leq k$, we define $\I_i$ in an iterative process starting from an
  arbitrary interpretation $\I_0$ stemming from \cW and setting
  \[
  \I_i := \I_{i-1}[A_i \mapsto \{ x \in \Delta^\cW \mid x \in D_i^{\I_{i-1}}
  \}]
  \]
  
  Since, for each $A_i$ there is exactly one axiom in \cT, each step in this
  process is well-defined. Also, since $\Int(\sL)$ may only restrict the
  interpretation of atomic roles, $\I_i \in \Int(\sL)$ for each $0 \leq i \leq
  k$. For $\I = \I_k$ it can be shown that \I is an interpretation stemming
  from \cW with $\I \models \cT$. 
  
  First we prove inductively that, for $0 \leq i \leq k$, $\I_i$ stems from \cW.
  We have already required $\I_0$ to stem from \cW.
  
  Assume the claim was proved for $\I_{i-1}$ and $\I_i$ does not stem from
  \cW. Then there must be some $x \in \Delta^\cW$ such that either (i) $A_i
  \in \LabW(x)$ but $x \not \in A_i^{\I_i}$ or (ii) $\neg A_i \in \LabW(x)$
  but $x \in A_i^{\I_i}$ (since we assume $\I_{i-1}$ to stem from \cW and
  $A_i$ is the only atomic concept whose interpretation changes from
  $\I_{i-1}$ to $\I_i$). The two cases can be handled dually:
  \begin{itemize}
  \item[(i)] From $A_i \in \LabW(x)$ it follows that $D_i \in \LabW(x)$,
    because \cW is unfolded. Since $\I_{i-1}$ stems from \cW and \cW is a
    witness, Property \witn 3 implies $x \in D_i^{\I_{i-1}}$. But this
    implies $x \in A_i^{\I_i}$, which is a contradiction.
  \item[(ii)] From $\neg A_i \in \LabW(x)$ it follows that $\neg D_i \in
    \LabW(x)$ because \cW is unfolded. Since $\I_{i-1}$ stems from \cW and
    \cW is an witness, Property \witn 3 implies $x \in (\neg
    D_i)^{\I_{i-1}}$. Since $(\neg D_i)^{\I_{i-1}} = \Delta^\cW \setminus
    D_i^{\I_{i-1}}$ this implies $x \not\in A_i^{\I_i}$, which is a
    contradiction.
  \end{itemize}
  Together this implies that $\I_i$ also stems from \cW.
  
  To show that $\I \models \cT$ we show inductively that $\I_i \models A_j
  \doteq D_j$ for each $1 \leq j \leq i$. This is obviously true for $i = 0$.
  
  The interpretation of $D_i$ may not depend on the interpretation of
  $A_i$ because otherwise \interp 2 would imply that $A_i$ uses
  itself. Hence $D_i^{\I_i} = D_i^{\I_{i-1}}$ and,  by
  construction, $\I_i \models A_i \doteq D_i$. Assume there is
  some $j < i$ such that $\I_i \not\models A_j \doteq D_j$. Since $\I_{i-1}
  \models A_j \doteq D_j$ and only the interpretation of $A_i$ has changed
  from $\I_{i-1}$ to $\I_i$, $D_j^{\I_i} \neq D_j^{\I_{i-1}}$ must hold
  because of \interp 2. But
  this implies that $A_i$ occurs in $D_j$ and hence $A_j$ uses $A_i$
  which contradicts $j < i$.  Thus, we have $\I \models A_j = D_j$ for each
  $1 \leq j \leq k$ and hence $\I \models \cT$. \qed
\end{proof}

Lazy unfolding is a well-known and widely used technique for optimising
reasoning \wrt primitive TBoxes~\cite{BFHNP94}. So far, we have only given a correctness
proof for this relatively simple approach, although one that is independent of a specific \DL or
reasoning algorithm. With the next lemma we show how we can extend correct
absorptions and hence how lazy unfolding can be applied to 
a broader class of TBoxes. A further enhancement of the technique is presented
in Section~\ref{sec:performance}.

\begin{lemma}\label{lem:extension-lemma}
  Let  $(\Tu,\Tg)$ be a correct absorption of a TBox 
  \cT. 
  \begin{enumerate}
  \item If $\cT'$ is an arbitrary TBox, then $(\Tu,\Tg \cup \cT')$ is a
    correct absorption of $\cT \cup \cT'$.
  \item If $\cT'$ is a TBox that consists entirely of axioms of the form $A
    \sqsubseteq D$, where $A \in \NC$ and $A$ is not defined in $\Tu$, then
    $(\Tu \cup \cT',\Tg)$ is a correct absorption of $\cT \cup \cT'$.
  \end{enumerate}
\end{lemma}

\begin{proof} In both cases, $\Tu \cup \Tg \cup \cT' \equiv \cT \cup \cT'$ holds trivially.
  \begin{enumerate}
  \item Let $C \in \sL$ be a concept and $\cW$ be an unfolded witness for $C$
    \wrt the absorption $(\Tu,\Tg \cup \cT')$. This implies that \cW is unfolded
    \wrt the (smaller) absorption $(\Tu,\Tg)$. Since $(\Tu,\Tg)$ is a correct
    absorption, there is an interpretation $\I$ stemming from \cW with $\I
    \models \cT$. Assume $\I \not \models \cT'$. Then, without loss of
    generality,\footnote{Arbitrary TBoxes can be expressed using only
      axioms of the form $C \sqsubseteq D$.} there is an axiom $D
    \sqsubseteq E \in \cT'$ such that there exists an $x \in D^\I \setminus
    E^\I$. Since \cW is unfolded, we have $D \rightarrow E \in \Lab^\cW(x)$
    and hence \witn 3 implies $x \in (\neg D \sqcup E)^\I =\Delta^\I \setminus
    (D^\I \setminus E^\I)$, a contradiction. Hence $\I \models \cT \cup \cT'$
    and \cW is admissible \wrt $\cT \cup \cT'$.
  \item Let $C \in \sL$ be a concept and $\cW$ be an unfolded witness for $C$
    w.r.t.\ the absorption $(\Tu \cup \cT',\Tg)$. From $\cW$ we define a new witness
    $\cW'$ for $C$ by setting $\Delta^{\cW'}  := \Delta^\cW$, $\cdot^{\cW'}
    := \cdot^\cW$, and definig $\Lab^{\cW'}$ to be the function that, for
    every $x \in \Delta^{\cW'}$, maps $x$ to the set
    \[
    \Lab^\cW(x) \cup \{ \neg A \mid A \sqsubseteq
    D \in \cT', A \not \in \Lab^\cW(x) \}
    \]
    
    It is easy to see that $\cW'$ is indeed a witness for $C$ and that $\cW'$
    is also unfolded w.r.t.\ the absorption $(\Tu \cup \cT',\Tg)$. This
    implies that $\cW'$ is also unfolded w.r.t.\ the (smaller) absorption
    $(\Tu,\Tg)$.  Since $(\Tu,\Tg)$ is a correct absorption of $\cT$, there
    exists an interpretation \I stemming from $\cW'$ such that $\I \models
    \cT$.  We will show that $\I \models \cT'$ also holds. Assume $\I \not
    \models \cT'$, then there is an axiom $A \sqsubseteq D \in \cT'$ and an $x
    \in \Delta^\I$ such that $x \in A^\I$ but $x \not \in D^\I$. By
    construction of $\cW'$, $x \in A^\I$ implies $A \in \Lab^{\cW'}(x)$
    because otherwise $\neg A \in \Lab^{\cW'}(x)$ would hold in contradiction
    to \witn 3. Then, since $\cW'$ is unfolded, $D \in \Lab^{\cW'}(x)$, which,
    again by \witn 3, implies $x \in D^\I$, a contradiction.
  
    Hence, we have shown that there exists an interpretation $\I$ stemming
    from $\cW'$ such that $\I \models \Tu \cup \cT' \cup \Tg$. By construction
    of $\cW'$, any interpretation stemming from $\cW'$ also stems from $\cW$,
    hence $\cW$ is admissible \wrt $\cT \cup \cT'$. \qed
\end{enumerate}
\end{proof}

\section{APPLICATION TO \Fact}
\label{sec:application}

\newcommand{\Tprim}{\ensuremath{\cT_{\text{prim}}}\xspace}
\newcommand{\Trest}{\ensuremath{\cT_{\text{inc}}}\xspace}

In the preceeding section we have defined correct absorptions and discussed
how they can be exploited in order to optimise satisfiability procedures. However, we have said
nothing about the problem of how to find an absorption given an arbitrary
terminology. In this section we will describe the absorption algorithm used by
\Fact and prove that it generates correct absorptions.

Given a TBox $\cT$ containing arbitrary axioms, the absorption algorithm used
by \Fact constructs a triple of TBoxes $(\Tg,\Tprim,\Trest)$ such that
\begin{itemize}
\item $\cT \equiv \Tg \cup \Tprim \cup \Trest$,
\item $\Tprim$ is primitive, and
\item $\Trest$ consists only of axioms of the form $A \sqsubseteq D$ where $A
  \in \NC$ and $A$ is not defined in $\Tprim$.
\end{itemize}
We refer to these properties by $(*)$.  From Theorem~\ref{theo:prim-tbox}
together with  Lemma~\ref{lem:extension-lemma} it follows that,
for
\[
\Tu := \{ A \sqsubseteq D, \neg A \sqsubseteq \neg D \mid A \doteq D \in
\Tprim \} \cup \Trest
\]
(\Tu,\Tg) is a correct absorption of $\cT$; hence satisfiability for a
concept $C$ \wrt \cT can be decided by checking for an unfolded witness for $C$.

In a first step, \Fact distributes axioms from \Tau amongst $\Trest$,
$\Tprim$, and $\Tg$, trying to minimise the number of axioms in $\Tg$
while still maintaining $(*)$. To do this, it initialises $\Tprim,
\Trest$, and $\Tg$ with $\emptyset$, and then processes each axiom $X
\in \Tau$ as follows.
\begin{enumerate}
\item If $X$ is of the form $A \sqsubseteq C$, then
  \begin{enumerate}
  \item if $A \in \NC$ and $A$ is not defined in $\Tprim$ then $X$
    is added to $\Trest$,
  \item otherwise $X$ is added to $\Tg$
  \end{enumerate}
\item If $X$ is of the form $A \doteq C$, then
  \begin{enumerate}
  \item if $A \in \NC$, $A$ is not defined in 
    $\Tprim$ or $\Trest$ and  $\Tprim \cup \{X\}$ is primitive, then $X$ is added
    to $\Tprim$,
  \item otherwise, the axioms $A \sqsubseteq C$ and $C \sqsubseteq A$
    are added to $\Tg$.
  \end{enumerate}
\item If $X$ is of the form $C \sqsubseteq D$, then add $X$ to $\Tg$
\item If $X$ is of the form $C \doteq D$, then add $C \sqsubseteq D, D
  \sqsubseteq C$ to $\Tg$.
\end{enumerate}

It is easy to see that the resulting TBoxes $\Tg, \Tprim, \Trest$ satisfy
$(*)$.  In a second step, \Fact processes the axioms in $\Tg$ one at a
time, trying to absorb them into axioms in $\Trest$.  Those axioms
that are not absorbed remain in $\Tg$.
To
give a simpler formulation of the algorithm, each axiom $(C \sqsubseteq D) \in
\Tg$ is viewed as a clause $\mathbf{G}=\{D, \Not C\}$, corresponding to
the axiom $\top \sqsubseteq C \rightarrow D$, which is equivalent to $C
\sqsubseteq D$. For each such axiom \Fact applies the following absorption procedure.
\begin{enumerate}
\item \label{09-item:abs2} Try to absorb $\mathbf{G}$. If there is a concept
  $\neg A \in \mathbf{G}$ such that $A \in \NC$ and $A$ is not defined in
  $\Tprim$, then add $A \sqsubseteq B$ to $\Trest$, where $B$ is the
  disjunction of all the concepts in $\mathbf{G} \setminus \{\neg A\}$, remove
  $\mathbf{G}$ from $\Tg$, and exit.
\item  Try to simplify $\mathbf{G}$.
  \begin{enumerate}
  \item If there is some $\Not C \in \mathbf{G}$ such that $C$ is of
    the form $C_1 \sqcap \ldots
    \sqcap C_n$, then substitute $\Not C$ with $\Not C_1 \sqcup \ldots
    \sqcup \Not C_n$, and %return to step~\ref{09-item:abs2}.
    continue with step~\ref{09-item:abs2b}.
  \item \label{09-item:abs2b} If there is some $C \in \mathbf{G}$ such that $C$ is of the
    form $(C_1 \sqcup \ldots \sqcup C_n)$, then apply associativity by
    setting $\mathbf{G} = \mathbf{G} \cup \{C_1, \ldots, C_n\}
    \setminus \{(C_1 \sqcup \ldots \sqcup C_n)\}$, and return to
    step~\ref{09-item:abs2}.
  \end{enumerate}
\item Try to unfold $\mathbf{G}$. If, for some $A \in \mathbf{G}$
  (resp.\ $\neg A \in \mathbf{G}$), there is an axiom $A \doteq C$ in
  $\Tprim$, then substitute $A \in \mathbf{G}$ (resp.\ $\neg A \in
  \mathbf{G}$) with $C$ (resp.\ $\neg C$) and return to
  step~\ref{09-item:abs2}.
\item If none of the above were possible, then absorption of $\mathbf{G}$ 
  has failed. Leave $\mathbf{G}$ in $\Tg$, and exit.
\end{enumerate}

For each step, we have to show that $(*)$ is maintained. Dealing with clauses
instead of axioms causes no problems. In the first step, axioms are moved
from $\Tg$ to $\Trest$ as long as this does not violate $(*)$.
The second and the third step replace a clause by an equivalent one and hence
do not violate $(*)$.

Termination of the procedure is obvious. Each axiom is considered only
once and, for a given axiom, simplification and unfolding can only be
applied finitely often before the procedure is exited, either by
absorbing the axiom into $\Trest$ or leaving it in $\Tg$. For
simplification, this is obvious; for unfolding, this holds because
$\Tprim$ is acyclic. Hence, we get the following:

\begin{theorem}
  For any TBox \Tau, \Fact computes a correct absorption of \Tau.
\end{theorem}

\section{IMPROVING PERFORMANCE}
\label{sec:performance}

\newcommand{\dotcup}{\mathbin{\dot \cup}}
\newcommand{\shiq}{\ensuremath{\mathcal{SHIQ}}\xspace}

\begin{figure*}[htb]
  \begin{center}
        \includegraphics[width=0.75\linewidth]{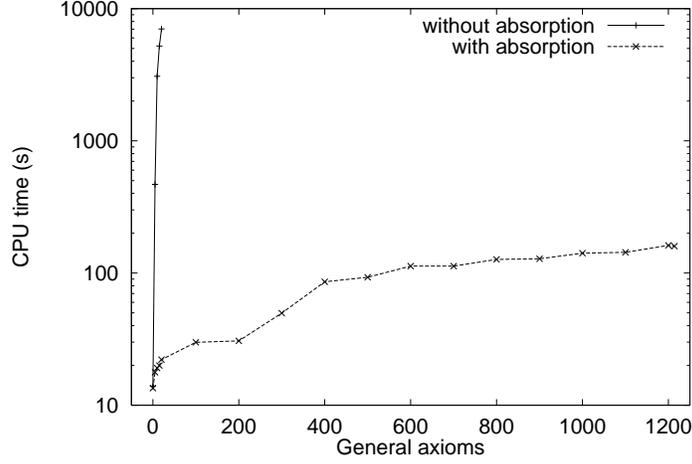}
  \end{center}
  \caption{Classification times with and without absorption}
  \label{fig:graph1}
\end{figure*}

The absorption algorithm employed by \Fact already leads to a dramatic
improvement in performance. This is illustrated by
Figure~\ref{fig:graph1}, which shows the times taken by \Fact to
classify versions of the \Galen KB with some or all of the general
axioms removed. Without absorption, classification time increased
rapidly with the number of general axioms, and exceeded 10,000s with
only 25 general axioms in the KB; with absorption, only 160s was taken
to classify the KB with all 1,214 general axioms.

However, there is still considerable scope
for further gains. In particular, the following definition for a
\emph{stratified} TBox allows lazy unfolding to be more generally
applied, while still allowing for correct absorptions.

\begin{definition}[Stratified TBox]\label{def:monotone-tbox}
  A TBox \cT is called \emph{stratified} iff it consists entirely of axioms of the form
  $A \doteq D$ with $A \in \NC$, each $A \in \NC$ appears at most once
  on the left-hand side of an axiom, and \cT can be arranged
  monotonously, i.e., there is a 
  disjoint partition  $\cT_1 \dotcup \cT_2 \dotcup \dots \dotcup \cT_k$ of $\cT$, such that
  \begin{itemize}
  \item for all $1 \leq j < i \leq k$, if $A \in \NC$ is defined in $\cT_i$,
    then it does not occur in $\cT_j$, and
  \item for all $1\leq i \leq k$, all concepts which appear on the right-hand
    side of axioms in $\cT_i$ are monotone in all atomic concepts defined in
    $\cT_i$. 
  \end{itemize}
  A concept $C$ is monotone in an atomic concept $A$ if, for any
  interpretation $\I \in \Int(\sL)$ and any two sets $X_1,X_2 \subseteq
  \Delta^\I$,
  \[
  X_1 \subseteq X_2 \ \Implies \  C^{\I[A \mapsto X_1]} \subseteq
  C^{\I[A \mapsto X_2]} .
  \]
\end{definition}

For many DLs, a sufficient condition for monotonicity is \emph{syntactic}
monotonicity, i.e., a concept $C$ is syntactically monotone in some atomic
concept $A$ if $A$ does no appear in $C$ in the scope of an odd number of
negations.

Obviously, due to its acyclicity, every primitive TBox is also stratified and
hence the following theorem is a strict generalisation of
Theorem~\ref{theo:prim-tbox}.

\begin{theorem}\label{theo:monotone-tbox}
  Let \cT be a stratified TBox, $\Tg = \emptyset$ and \Tu defined by
  \[
  \Tu = \{ A \sqsubseteq D, \neg A \sqsubseteq \neg D \mid A \doteq D \in \cT
  \} .
  \]
  Then $(\Tu,\Tg)$ is a correct absorption of \cT.
\end{theorem}

The proof of this theorem follows the same line as the proof of
Theorem~\ref{theo:prim-tbox}. Starting from an arbitrary interpretation $\I_0$
stemming from the unfolded witness, we incrementally construct interpretations
$\I_1, \dots, \I_k$, using a fixed point construction in each step. We show
that each $\I_i$ stems from \cW and that, for $1 \leq j < i \leq k$, $\I_i
\models \cT_j$, hence $\I_k \models \cT$ and stems from $\cW$.

Before we prove this theorem, we recall some basics of lattice theory.
For any set $\mathcal{S}$, the powerset of $\mathcal{S}$, denoted by
$2^{\mathcal{S}}$ forms a complete lattice, where the ordering, join and meet
operations are set-inclusion $\subseteq$, union $\cup$, and intersection
$\cap$, respectively. For any complete lattice $\mathcal{L}$, its $n$-fold
cartesian product $\mathcal{L}^n$ is also a complete lattice, with ordering,
join, and meet defined in a pointwise manner.

For a lattice $\mathcal{L}$, a function $\Phi : \mathcal{L} \rightarrow
\mathcal{L}$ is called monotone, iff, for $x_1,x_2 \in \mathcal{L}$, $x_1
\sqsubseteq x_2$ implies $\Phi(x_1) \sqsubseteq \Phi(x_2)$.

By Tarski's fixed point theorem \cite{Tars55}, every monotone function $\Phi$
on a complete lattice, has uniquely defined least and greatest fixed points,
i.e., there are elements $\overline x, \underline x \in \mathcal{L}$ such that
\[
\overline x = \Phi(\overline x) \textsf{ and } \underline x = \Phi(\underline x)
\]
and, for all $x \in \mathcal{L}$ with $x = \Phi(x)$, 
\[
\underline x \sqsubseteq x \textsf{ and } x \sqsubseteq \overline x .
\]

\noindent \textit{Proof of Theorem~\ref{theo:monotone-tbox}.}
  $\Tu \cup \Tg \equiv \cT$ is obvious. Let $\cW =
  (\Delta^\cW,\cdot^\cW,\LabW)$ be an unfolded witness.  We have to show that
  there is an interpretation \I stemming from \cW with $\I \models \cT$. Let
  $\cT_1, \dots, \cT_k$ be the required partition of $\cT$. We will define \I
  inductively, starting with an arbitrary interpretation $\I_0$ stemming from
  \cW.

  Assume $\I_{i-1}$ was already defined. We define $\I_i$ from $\I_{i-1}$ as
  follows: let $\{ A^i_1 \doteq D^i_1, \dots, A^i_m \doteq D^i_m \}$ be an
  enumeration of  $\cT_i$. First we need some auxiliary notation: for any
  concept $C \in \sL$ we define
  \[
  C^\cW := \{ x \in \Delta^\cW \mid C \in \LabW(x) \} .
  \]
  
  Using this notation we define the function $\Phi$ mapping subsets
  $X_1,\dots,X_m$ of $\Delta^\cW$ to
  \begin{gather*}
    \begin{array}{l@{\;}l@{\;}l}
      ( & ( (A^i_1)^\cW
      %\{x \mid A^i_1 \in \Lab^\cW(x) \} 
      \cup
      (D^i_1)^{\I_{i-1}(X_1,\dots,X_m)} ) \setminus 
      (\neg A^i_1)^\cW, \\
      %\{ x \mid \neg A^i_1 \in  \Lab^\cW(x) \}, \\
      & \dots, \\
      & ( (A^i_m)^\cW
      % \{x \mid A^i_m \in \Lab^\cW(x) \} 
      \cup
      (D^i_m)^{\I_{i-1}(X_1,\dots,X_m)})  \setminus 
      (\neg A^i_m)^\cW
      %\{ x \mid \neg A^i_m \in \Lab^\cW(x) \}  
      &) 
    \end{array}\\
    \intertext{where}
    \I_{i-1}(X_1,\dots,X_m) := \I_{i-1}[A^i_1 \mapsto X_1, \dots, A^i_m
    \mapsto X_m]
  \end{gather*}
  Since all of the $D^i_j$ are monotone in all of the $A^i_m$, $\Phi$ is a
  monontone function. This implies that $\Phi$ has a least fixed point, which
  we denote by $(\underline X_1, \dots, \underline X_m)$. We use this fixed
  point to define $\I_i$ by
  \[
  \I_i := \I_{i-1}[A^i_1 \mapsto \underline X_1, \dots, A^i_m \mapsto \underline X_m]
  \]

  \noindent \textsc{Claim 1:} For each $0 \leq i \leq k$, $\I_i$ stems
  from \cW. 

  We show this claim by induction on $i$. We have already required $\I_0$ to
  stem from $\cW$.  Assume $\I_{i-1}$ stems from $\cW$. Since the only thing that
  changes from $\I_{i-1}$ to $\I_i$ is the interpretation of the atomic
  concepts $A^i_1, \dots, A^i_m$, we only have to check that $A^i_j \in
  \Lab^\cW(x)$ implies $x \in (A^i_j)^{\I_i}$ and $\neg A^i_j \in \Lab^\cW(x)$
  implies $x \not \in (A^i_j)^{\I_i}$.
  
  By definition of $\Phi$, and because $\{x \mid A^i_j \in \Lab^\cW(x) \} \cap
  \{x \mid \neg A^i_j \in \Lab^\cW(x) \} = \emptyset$, $A^i_j \in \Lab^\cW(x)$
  implies $x \in (A^i_j)^{\I_i}$.  Also by the definition of $\Phi$,
  $\neg A^i_j \in \Lab^\cW(x)$ implies $x \not \in (A^i_j)^{\I_i}$. Hence,
  $\I_i$ stems from $\cW$.

  \noindent \textsc{Claim 2:} For each $1 \leq j \leq i \leq k$, $\I_i \models \cT_j$.

  We prove this claim by induction over $i$ starting from $0$. For $i=0$,
  there is nothing to prove. Assume the claim would hold for $\I_{i-1}$. The
  only thing that changes from $\I_{i-1}$ to $\I_i$ is the interpretation of
  the atomic concepts $A^i_1,\dots A^i_m$ defined in $\cT_i$. Since these
  concepts may not occur in $\cT_j$ for $j < i$, the interpretation of the
  concepts in these TBoxes does not change, and from $\I_{i-1} \models \cT_j$
  follows $\I_i \models \cT_j$ for $1 \leq j \leq i-1$.

  It remains to show that $\I_i \models \cT_i$. Let $A^i_j \doteq D^i_j$ be an 
  axiom from $\cT_i$. From the definition of $\I_i$ we have 
  \begin{equation}
    \label{eq:eq1}
    (A^i_j)^{\I_i} = ( (A^i_j)^\cW 
    % \{ x \mid A^i_j \in \Lab^\cW(x) \}
    \cup (D^i_j)^{\I_i} )
    \setminus 
    (\neg A^i_j)^\cW .
    % \{ x \mid \neg A^i_j \in \Lab^\cW(x) \} .
  \end{equation}
  
  $\cW$ is unfolded, hence $A^i_j \in \Lab^\cW(x)$ implies $D^i_j \in
  \Lab^\cW(x)$ and, since $\I_i$ stems from $\cW$, this implies $x \in
  (D^i_j)^{\I_i}$, thus
  \begin{equation}
    \label{eq:eq2}
    (A^i_j)^\cW 
    % \{ x \mid A^i_j \in \Lab^\cW(x) \} 
    \cup (D^i_j)^{\I_i} = (D^i_j)^{\I_i}
  \end{equation}
  Furthermore, $\neg A^i_j \in \Lab^\cW(x)$ implies $\neg D^i_j \in
  \Lab^\cW(x)$ implies $x \in (\neg D^i_j)^{\I_i}$, thus
  \begin{equation}
    \label{eq:eq3}
    (D^i_j)^{\I_i} \setminus 
    (\neg A^i_j)^\cW 
    % \{ x \mid \neg A^i_j \in \Lab^\cW(x) \} 
    = (D^i_j)^{\I_i}
  \end{equation}
  
  Taking together (\ref{eq:eq1}), (\ref{eq:eq2}), and (\ref{eq:eq3}) we get
  \[
  (A^i_j)^{\I_i} = (D^i_j)^{\I_i} ,
  \]
  and hence $\I_i \models A^i_j \doteq D^i_j$.
  
  Together, Claim 1 and Claim 2 prove the theorem, since $\I_k$ is an
  interpretation that stems from $\cW$ and satisfies $\cT$. \qed

This theorem makes it possible to apply the same lazy unfolding strategy as
before to cyclical definitions. Such definitions are quite natural in a logic
that supports inverse roles. For example, an orthopaedic procedure might be
defined as a procedure performed by an orthopaedic surgeon, while an
orthopaedic surgeon might be defined as a surgeon who performs only
orthopaedic procedures:\footnote{This example is only intended for didactic
  purposes.}
\[
\begin{array}{@{}r@{\;}c@{\;}l@{}}
\Cname{o-procedure} & \doteq & \Cname{procedure} \sqcap (\Some{\Rname{performs}^-}{\Cname{o-surgeon}})\\
\Cname{o-surgeon} & \doteq & \Cname{surgeon} \sqcap
(\All{\Rname{performs}}{\Cname{o-procedure}})
\end{array}
\]

The absorption algorithm described in Section~\ref{sec:application}
would force the second of these definitions to be added to $\Tg$ as
two general axioms and, although both axioms would subsequently be
absorbed into $\Tu$, the procedure would result in a disjunctive term
being added to one of the definitions in $\Tu$. Using
Theorem~\ref{theo:monotone-tbox} to enhance the absorption algorithm
so that these kinds of definition are directly added to $\Tu$ reduces
the number of disjunctive terms in $\Tu$ and can lead to significant
improvements in performance.

\begin{figure*}[htb]
  \begin{center}
        \includegraphics[width=0.75\linewidth]{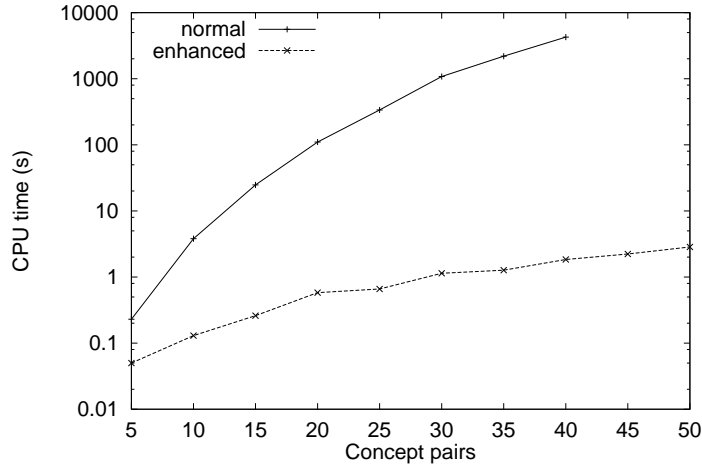}
  \end{center}
  \caption{Classification times with and without enhanced absorption}
  \label{fig:graph2}
\end{figure*}

This can be demonstrated by a simple experiment with the new \Fact
system, which implements the \shiq logic~\cite{Horrocks99j} and is
thus able to deal with inverse roles. Figure~\ref{fig:graph2} shows the
classification time in seconds using the normal and enhanced
absorption algorithms for terminologies consisting of between 5 and 50
pairs of cyclical definitions like those described above for
\Cname{o-surgeon} and \Cname{o-procedure}. With
only 10 pairs the gain in performance is already a factor of 30, while
for 45 and 50 pairs it has reached several orders of magnitude: with
the enhanced absorption the terminology is classified in 2--3
seconds whereas with the original algorithm the time required exceeded
the 10,000 second limit imposed in the experiment.

It is worth pointing out that it is by no means trivially true that
cyclical definitions can be dealt with by lazy unfolding. Even
without inverse roles it is clear that definitions such as $A \doteq
\Not A$ (or more subtle variants) force the domain to be empty and
would lead to an incorrect absorption if dealt with by lazy unfolding.
With converse roles it is, for example, possible to force the
interpretation of a role $R$ to be empty with a definition such as $A
\doteq \All{R}{(\All{R^-}{\Not A})}$, again leading to an incorrect
absorption if dealt with by lazy unfolding.

\section{OPTIMAL ABSORPTIONS}

We have demonstrated that absorption is a highly effective and widely
applicable technique, and by formally defining correctness criteria for
absorptions we have proved that the procedure used by \Fact finds
correct absorptions. Moreover, by establishing more precise correctness
criteria we have demonstrated how the effectiveness of this procedure
could be further enhanced.

However, the absorption algorithm used by \Fact is clearly
sub-optimal, in the sense that changes could be made that would, in
general, allow more axioms to be absorbed (e.g., by also giving
special consideration to axioms of the form $\Not A \sqsubseteq C$
with $A \in \NC$). Moreover, the procedure is non-deterministic, and,
while it is guaranteed to produce a correct absorption, its specific
result depends on the order of the axioms in the original TBox \Tau.
Since the semantics of a TBox \Tau does not depend on the order of its
axioms, there is no reason to suppose that they will be arranged in a
way that yields a ``good'' absorption. Given the 
effectiveness of absorption, it would be desirable to have an
algorithm that was guaranteed to find the ``best'' absorption possible for
any set of axioms, irrespective of their ordering in the TBox.

Unfortunately, it is not even clear how to define a sensible
optimality criterion for absorptions. It is obvious that simplistic
approaches based on the number or size of axioms remaining in $\Tg$
will not lead to a useful solution for this problem. Consider, for
example, the cyclical TBox experiment from the previous section. Both the  original
\Fact absorption algorithm and the enhanced algorithm, which exploits
Theorem~\ref{theo:monotone-tbox}, are able to compute a complete absorption of
the axioms ( i.e., a correct absorption with $\Tg=\emptyset$), but
the enhanced algorithm leads to much better performance, as
shown in Figure~\ref{fig:graph2}.

An important issue for future work is, therefore, the identification
of a suitable optimality criterion for absorptions, and the
development of an algorithm that is able to compute absorptions that
are optimal with respect to this criterion.

\subsubsection*{Acknowledgements}

This work was partially supported by the DFG, Project No. GR 1324/3-1.

\newcommand{\etalchar}[1]{$^{#1}$}


\begin{thebibliography}{BHH{\etalchar{+}}91}

\bibitem[Baa91]{Baader91c}
F.~Baader.
\newblock Augmenting concept languages by transitive closure of roles: An
  alternative to terminological cycles.
\newblock In {\em Proceedings of the 12th International Joint Conference on
  Artificial Intelligence (IJCAI-91)}, pages 446--451, 1991.

\bibitem[BBH96]{BaaderBuchheit+-AIJ-1996}
F.~Baader, M.~Buchheit, and B.~Hollunder.
\newblock Cardinality restrictions on concepts.
\newblock {\em Artificial Intelligence}, 88(1--2):195--213, 1996.

\bibitem[BDS93]{Buchheit93}
M.~Buchheit, F.~M. Donini, and A.~Schaerf.
\newblock Decidable reasoning in terminological knowledge representation
  systems.
\newblock {\em J.\ of Artificial Intelligence Research}, 1:109--138, 1993.

\bibitem[BFH{\etalchar{+}}94]{BFHNP94}
F.~Baader, E.~Franconi, B.~Hollunder, B.~Nebel, and H.-J. Profitlich.
\newblock An empirical analysis of optimization techniques for terminological
  representation systems, or: Making {KRIS} get a move on.
\newblock {\em Applied Artificial Intelligence}, 4:109--132, 1994.

\bibitem[BFH{\etalchar{+}}99]{Horrocks99i}
A.~Borgida, E.~Franconi, I.~Horrocks, D.~McGuinness, and P.~F. Patel-Schneider.
\newblock Explaining $\mathcal{ALC}$ subsumption.
\newblock In P.~Lambrix, A.~Borgida, M.~Lenzerini, R.~M{\"o}ller, and
  P.~Patel-Schneider, editors, {\em Proceedings of the International Workshop
  on Description Logics (DL'99)}, pages 37--40, 1999.

\bibitem[BH91]{Baader91e}
F.~Baader and B.~Hollunder.
\newblock A terminological knowledge representation system with complete
  inference algorithms.
\newblock In {\em Processing declarative knowledge: International workshop
  PDK'91}, number 567 in Lecture Notes in Artificial Intelligence, pages
  67--86, Berlin, 1991. Springer-Verlag.

\bibitem[BHH{\etalchar{+}}91]{Baader91b}
F.~Baader, H.-J. Heinsohn, B.~Hollunder, J.~Muller, B.~Nebel, W.~Nutt, and
  H.-J. Profitlich.
\newblock Terminological knowledge representation: A proposal for a
  terminological logic.
\newblock Technical Memo TM-90-04, Deutsches Forschungszentrum f{\"u}r
  K{\"u}nstliche Intelligenz GmbH (DFKI), 1991.

\bibitem[Cal96]{Calvanese96a}
D.~Calvanese.
\newblock Reasoning with inclusion axioms in description logics: Algorithms and
  complexity.
\newblock In Wolfgang Wahlster, editor, {\em Proceedings of the 12th European
  Conference on Artificial Intelligence (ECAI'96)}, pages 303--307. John Wiley
  \& Sons Ltd., 1996.

\bibitem[CDL99]{CaDL99}
D.~Calvanese, G.~{De Giacomo}, and M.~Lenzerini.
\newblock Reasoning in expressive description logics with fixpoints based on
  automata on infinite trees.
\newblock In {\em Proc.\ of the 16th Int.\ Joint Conf.\ on Artificial
  Intelligence (IJCAI'99)}, 1999.

\bibitem[DDM96]{Donini96a}
F.~Donini, G.~{De~Giacomo}, and F.~Massacci.
\newblock {EXPTIME} tableaux for $\mathcal{ALC}$.
\newblock In L.~Padgham, E.~Franconi, M.~Gehrke, D.~L. McGuinness, and P.~F.
  Patel-Schneider, editors, {\em Collected Papers from the International
  Description Logics Workshop (DL'96)}, number WS-96-05 in AAAI Technical
  Report, pages 107--110. AAAI Press, Menlo Park, California, 1996.

\bibitem[DL94]{DeLe94}
G.~{De Giacomo} and M.~Lenzerini.
\newblock Boosting the correspondence between description logics and
  propositional dynamic logics.
\newblock In {\em Proc.\ of the 12th Nat.\ Conf.\ on Artificial Intelligence
  (AAAI'94)}, pages 205--212. {AAAI} Press/The {MIT} Press, 1994.

\bibitem[DL96]{DeLe96}
G.~{De Giacomo} and M.~Lenzerini.
\newblock {TBox} and {ABox} reasoning in expressive description logics.
\newblock In Luigia~C. Aiello, John Doyle, and Stuart~C. Shapiro, editors, {\em
  Proc.\ of the 5th Int.\ Conf.\ on the Principles of Knowledge Representation
  and Reasoning (KR'96)}, pages 316--327. Morgan Kaufmann, Los Altos, 1996.

\bibitem[DMar]{DeGiacomo98a}
G.~{{De~Giacomo}} and F.~Massacci.
\newblock Combining deduction and model checking into tableaux and algorithms
  for converse-{PDL}.
\newblock {\em Information and Computation: special issue on the Federated
  Logic Conferences}, to appear.

\bibitem[EH85]{Emerson85}
E.~A. Emerson and J.~Y. Halpern.
\newblock Decision procedures and expressiveness in the temporal logic of
  branching time.
\newblock {\em J.\ of Computer and System Sciences}, 30:1--24, 1985.

\bibitem[HN90]{Hollunder90b}
B.~Hollunder and W.~Nutt.
\newblock Subsumption algorithms for concept languages.
\newblock In {\em Proceedings of the 9th European Conference on Artificial
  Intelligence (ECAI'90)}, pages 348--353. John Wiley \& Sons Ltd., 1990.

\bibitem[HNS90]{HoNS90}
B.~Hollunder, W.~Nutt, and M.~{Schmidt-\hskip0pt Schauss}.
\newblock Subsumption algorithms for concept description languages.
\newblock In {\em ECAI-90}, Pitman Publishing, London, 1990.

\bibitem[Hor97]{Horrocks97b}
I.~Horrocks.
\newblock {\em Optimising Tableaux Decision Procedures for Description Logics}.
\newblock PhD thesis, University of Manchester, 1997.

\bibitem[Hor98]{Horrocks98c}
I.~Horrocks.
\newblock Using an expressive description logic: {FaCT} or fiction?
\newblock In A.~G. Cohn, L.~Schubert, and S.~C. Shapiro, editors, {\em
  Principles of Knowledge Representation and Reasoning: Proceedings of the
  Sixth International Conference (KR'98)}, pages 636--647. Morgan Kaufmann
  Publishers, San Francisco, California, June 1998.

\bibitem[HS99]{Hustadt99a}
U.~Hustadt and R.~A. Schmidt.
\newblock On the relation of resolution and tableaux proof systems for
  description logics.
\newblock In {\em Proceedings of the 16th International Joint Conference on
  Artificial Intelligence (IJCAI-99)}, pages 110--115, 1999.

\bibitem[HST99]{Horrocks99j}
I.~Horrocks, U.~Sattler, and S.~Tobies.
\newblock Practical reasoning for expressive description logics.
\newblock In {\em Proceedings of the 6th International Conference on Logic for
  Programming and Automated Reasoning (LPAR'99)}, pages 161--180, 1999.

\bibitem[RNG93]{Rector93a}
A.~L. Rector, W~A Nowlan, and A~Glowinski.
\newblock Goals for concept representation in the \textsc{Galen} project.
\newblock In {\em Proceedings of the 17th Annual Symposium on Computer
  Applications in Medical Care (SCAMC'93)}, pages 414--418, Washington DC, USA,
  1993.

\bibitem[Sch91]{Schi91}
K.~Schild.
\newblock A correspondence theory for terminological logics: Preliminary
  report.
\newblock In {\em Proc.\ of the 12th Int.\ Joint Conf.\ on Artificial
  Intelligence (IJCAI'91)}, pages 466--471, Sydney, 1991.

\bibitem[SS91]{SSSm91}
M.~{Schmidt-\hskip0pt Schau{\ss}} and G.~Smolka.
\newblock Attributive concept descriptions with complements.
\newblock {\em Acta Informatica}, 48(1):1--26, 1991.

\bibitem[Tar55]{Tars55}
A.~Tarski.
\newblock A lattice-theoretical fixpoint theorem and its applications.
\newblock {\em Pacific Journal of Mathematics}, 5:285--309, 1955.

\bibitem[VW86]{VaWo86}
M.~Y. Vardi and P.~Wolper.
\newblock Automata-theoretic techniques for modal logics of programs.
\newblock {\em J.\ of Computer and System Sciences}, 32:183--221, 1986.
\newblock A preliminary version appeared in {\sl Proc.\ of the 16th ACM SIGACT
  Symp.\ on Theory of Computing (STOC'84)}.

\bibitem[WS92]{Woods92}
W.~A. Woods and J.~G. Schmolze.
\newblock The \textsc{Kl-One} family.
\newblock {\em Computers and Mathematics with Applications -- Special Issue on
  Artificial Intelligence}, 23(2--5):133--177, 1992.

\end{thebibliography}
\end{document}